\documentclass[a4paper,11pt]{article}

\usepackage{contribution}
\newcommand{\weblink}[2][]{%
    \ifthenelse{\equal{#1}{}}%
    {\textnormal{\url{#2}}}%
    {\textnormal{\href{#2}{#1}}}%
}

\newcommand{\acknowledgements}[1]{%
  \bigskip\bigskip
  \textsf{\textbf{\Large Acknowledgements}} \\[2ex]
  {#1}
  \bigskip
}

\def\beq{\begin{equation}}
\def\eeq#1{\label{#1}\end{equation}}
\def\eeqn{\end{equation}}

\def\beqa{\begin{eqnarray}}
\def\eeqa#1{\label{#1}\end{eqnarray}}
\def\eeqan{\end{eqnarray}}

\let\bar=\overbar

\def\Dslash{\not{\hbox{\kern-4pt $D$}}}
\def\dslash{\not{\hbox{\kern-2pt $\del$}}}

\def\msb{{\bar{\ssstyle M \kern -1pt S}}}

\newcommand{\contribution}[7][]{%
  \clearpage
  \thispagestyle{plain}
  \ifthenelse{\equal{#1}{}}
  {\hypersetup{pdftitle={#2}}}
  {\hypersetup{pdftitle={#1}}}
  \hypersetup{pdfauthor={{#3} {#4}}}
  {\centering\normalfont\LARGE\bfseries\sffamily #2 \par\nobreak}
  \lhead{}
  \chead{%
    \textit{\footnotesize XIV International Conference on Hadron Spectroscopy
      (\weblink[\textit{hadron2011}]{http://www.hadron2011.de}), 13-17 June 2011, Munich, Germany}%
  }
  \rhead{}
  \bigskip
  \begin{center}
    {#3} {#4}\ifthenelse{\equal{#6}{}}{}{\footnote{\weblink[#6]{mailto:#6}}}
    \ifthenelse{\equal{#7}{}}{}{#7} \\
    \textit{#5}
  \end{center}
  \bigskip
}

\renewcommand{\abstract}[1]{%
  \begin{center}
    \begin{minipage}{0.85\textwidth}
      \begin{footnotesize}
        #1
      \end{footnotesize}
    \end{minipage}
  \end{center}
  \bigskip
}

\begin{document}

% % % % % % % % % % % % % % % % % % % % % % % % % % % % % % % % % % % % % % % % %
% your proceedings
%\input{contribution}

%%%%%%%%%%%%%%%%%%%%%%%%%%%%%%%%%%%%%%%%%%%%%%%%%%%%%%%%%%%%%%%%%%%%%%%%%%%%%%%%%
%
% template for hadron2011 contribution
%
% please do not rename this file
%
% to create document run
%
%     pdflatex hadron2011.tex
%
%%%%%%%%%%%%%%%%%%%%%%%%%%%%%%%%%%%%%%%%%%%%%%%%%%%%%%%%%%%%%%%%%%%%%%%%%%%%%%%%%
{  % do not remove

%%%%%%%%%%%%%%%%%%%%%%%%%%%%%%%%%%%%%%%%%%%%%%%%%%%%%%%%%%%%%%%%%%%%%%%%%%%%%%%%%
% please define your macros here

%
%%%%%%%%%%%%%%%%%%%%%%%%%%%%%%%%%%%%%%%%%%%%%%%%%%%%%%%%%%%%%%%%%%%%%%%%%%%%%%%%%

%%%%%%%%%%%%%%%%%%%%%%%%%%%%%%%%%%%%%%%%%%%%%%%%%%%%%%%%%%%%%%%%%%%%%%%%%%%%%%%%%
% define title, author, and address
% contribution[short title]{title}{author first name}{author last name}{author address}{author email}{collaboration}
% the short title will appear in the page headers and the TOC of the book of proceedings
% the last two arguments may be left empty
\contribution[Light Baryon Spectroscopy at CLAS]  % short title (optional)
{Light Baryon Spectroscopy using  the CLAS Spectrometer at Jefferson Laboratory}  % title
{Volker}{Crede}  % first and last name of author
{Department of Physics\\
  Florida State University\\
  Tallahassee, FL 32306, USA}  % author address
{crede@fsu.edu}  % author email optional
{on behalf of the CLAS Collaboration}  % collaboration (optional)
%
%%%%%%%%%%%%%%%%%%%%%%%%%%%%%%%%%%%%%%%%%%%%%%%%%%%%%%%%%%%%%%%%%%%%%%%%%%%%%%%%%

%%%%%%%%%%%%%%%%%%%%%%%%%%%%%%%%%%%%%%%%%%%%%%%%%%%%%%%%%%%%%%%%%%%%%%%%%%%%%%%%%
% abstract
\abstract{%
  Baryons are complex systems of confined quarks and gluons and
  exhibit the characteristic spectra of excited states. The systematics 
  of the baryon excitation spectrum is important to our understanding 
  of the effective degrees of freedom underlying nucleon matter.
  High-energy electrons and photons are a remarkably clean probe of
  hadronic matter, providing a microscope for examining the nucleon 
  and the strong nuclear force. Current experimental efforts with the
  CLAS spectrometer at Jefferson Laboratory utilize highly-polarized 
  frozen-spin targets in combination with polarized photon beams. 
  The status of the recent double-polarization experiments and some
  preliminary results are discussed in this contribution. 
}
%
%%%%%%%%%%%%%%%%%%%%%%%%%%%%%%%%%%%%%%%%%%%%%%%%%%%%%%%%%%%%%%%%%%%%%%%%%%%%%%%%%

%%%%%%%%%%%%%%%%%%%%%%%%%%%%%%%%%%%%%%%%%%%%%%%%%%%%%%%%%%%%%%%%%%%%%%%%%%%%%%%%%
% main text
% for short contributions sections are optional
\section{Introduction}

It is widely accepted that models based on three constituent-quark
degrees of freedom still provide the most comprehensive predictions of
the nucleon excitation spectrum. While many predicted properties of
the lower-mass (excited) states (< 1.8~GeV/$c^2$) agree fairly well
with experimental findings, discrepancies concerning the number
and ordering of states emerge above this threshold, mostly due to
missing experimental information. In recent years, lattice-QCD has 
made significant progress toward understanding the spectra of baryons,
despite the (still) large pion masses of about 420 MeV/$c^2$ used in
these calculations. Since baryon resonances are broad and overlapping,
individual excited states usually cannot be observed directly. To
extract resonance parameters, the observed angular distributions need
to be decomposed into partial waves in a partial wave analysis (PWA). 
Examples of PWA formalisms are described in~\cite{Dugger:2009pn,
Williams:2009rc}. Moreover, dynamical coupled channel models have 
been developed successfully in recent years from a more theoretical
side. The EBAC group at Jefferson Laboratory (JLab) has demonstrated
that the low physical mass of the Roper resonance can be explained by
such coupled channel effects~\cite{Kamano:2010ud}.

\subsection*{The Search for new Excited Baryons}
Differential cross sections alone result in ambiguous sets of resonances 
contributing to a particular photoproduction channel since almost
all information on interference effects is lost. For this reason, the
FROST experiment at JLab aims at performing so-called complete or
nearly-complete experiments for reactions like $\gamma p\to N\pi$,
$p\eta$, $p\omega$, $K^+Y$, and $p\pi^+\pi^-$, which will
significantly reduce and eventually eliminate the ambiguities in
the extraction of the scattering amplitude. The photoproduction of a
single pseudoscalar meson off the nucleon is fully described by four 
complex parity-conserving amplitudes, which may be determined from 
eight well-chosen combinations of the unpolarized cross section, three 
single-spin, and four double-spin observables~\cite{Chiang:1996em}.   

In the hyperon channels, precise cross section and polarization data
have been measured in recent years, e.g.~\cite{Bradford,Hleiqawi:2007ad,
AnefalosPereira:2009zw,McCracken:2009ra}.
The weak decay of the hyperon provides additional access to the
polarization of the recoiling hyperon rendering a complete experiment
feasible. If all combinations of beam, target, and recoil polarization
are measured, 16 observables can be extracted providing highly
redundant information on the production amplitude. In reactions
involving non-strange mesons (without measuring the recoil
polarization), seven independent observables can be directly
determined. The recoil polarization can then be inferred from
beam-target double-polarization measurements. In recent years, 
very precise differential cross section data were obtained for single
$p\pi^0,~n\pi^+,~p\eta,~p\eta^{\,\prime}$, and $p\omega$~production,
e.g. \cite{Dugger:2009pn,Chen:2009sda,Williams:2009rc}. Analyses on 
beam asymmetries for these reactions are currently being finalized. In
$\gamma p\to p\omega$, the $\omega$~decay to $\pi^+\pi^-\pi^0$ 
provides additional polarization information, which further constrains 
the partial wave analysis for this reaction~\cite{Williams:2009rc}. The
high-spin resonance, $N(2190)G_{17}$, decaying to $p\omega$ could 
be identified and confirmed in photoproduction as well as the weakly
established nucleon state, $N(1950)F_{15}$.

\section{Experimental Setup}
The results from the JLab double-polarization (FROST) measurements
discussed at this conference were obtained with the CEBAF Large
Acceptance Spectrometer (CLAS)~\cite{Mecking:2003zu} at the Thomas
Jefferson National Accelerator Facility. Longitudinally polarized
electrons with energies of 1.65 and 2.48~GeV were incident on the
thin radiator of the Hall B Photon Tagger~\cite{Sober:2000we} and
produced circularly-polarized tagged photons in the energy range
between 0.35 and 2.35~GeV with a polarization value of $\approx
85\,\%$ for the initial electron. The photon helicity was flipped at a
rate of 30~Hz. The frozen-spin butanol target had an average proton
spin state polarization of $\approx 82\,\%$ parallel to the beam axis
and $\approx 85\,\%$ anti-parallel to the beam axis. The average
target temperature was 30~mK with beam on target. Degradation of
target polarization occurred at rates of $\approx 0.9\,\%$ (parallel)
and $\approx 1.5\,\%$ (anti-parallel) per day. The target was typically 
repolarized once a week, usually with flips of the polarization
direction. Data were collected simultaneously for the butanol target
at the center of the CLAS detector, and, slightly downstream for
separate carbon and a polyethylene targets (to provide information
on bound nucleon backgrounds in the butanol target).

%%%%%%%%%%%%%%%%%%%%%%%%%%%%%%%%%%%%%%%%%%%%%%%%%%%%%%%%%%%%%%%%%%%%%%%%%%%%%%%%%
% the recommended way to include figures
%\begin{figure}[t]
%  \begin{center}
%    % please do not add file name extension this makes switching between latex and pdflatex easier
%    \includegraphics[width=1.0\textwidth]{example}
%    \caption{The magnet used in the Mesmeric studiesss.}
%    \label{fig:magnet}
%  \end{center}
%\end{figure}
%
%%%%%%%%%%%%%%%%%%%%%%%%%%%%%%%%%%%%%%%%%%%%%%%%%%%%%%%%%%%%%%%%%%%%%%%%%%%%%%%%%

%%%%%%%%%%%%%%%%%%%%%%%%%%%%%%%%%%%%%%%%%%%%%%%%%%%%%%%%%%%%%%%%%%%%%%%%%%%%%%%%%
% the recommended way to include a LaTeX table
%\begin{table}[tb]
%  \begin{center}
%    \begin{tabular}{lccc}  
%      Patient        & Initial level($\mu$g/cc) & w. Magnet & w. Magnet and Sound \\
%      \hline
%      Guglielmo B.   &                    0.12  &      0.10 &               0.001 \\
%      Ferrando di N. &                    0.15  &      0.11 &          $< 0.0005$ \\
%    \end{tabular}
%    \caption{Blood cyanide levels for the two patients.}
%    \label{tab:blood}
%  \end{center}
%\end{table}
%
%%%%%%%%%%%%%%%%%%%%%%%%%%%%%%%%%%%%%%%%%%%%%%%%%%%%%%%%%%%%%%%%%%%%%%%%%%%%%%%%%

\section{The Helicity Asymmetry $E$ for $\eta$ Photoproduction 
  on the Proton}

\begin{figure}[tb]
  \begin{center}
    % please do not add file name extension this makes switching between latex and pdflatex easier
    %\includegraphics[width=0.8\textwidth]{MORRISON/exciteV3}
    \epsfig{file=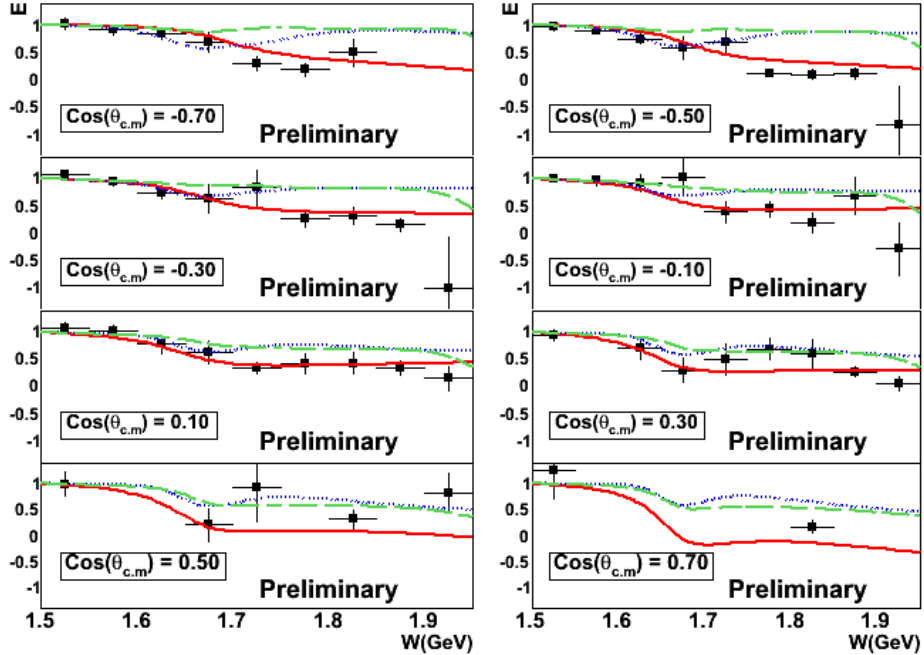,width=0.83\textwidth}
    \caption{Preliminary results for $E$ in $\vec{\gamma} \vec{p}\to
      p\eta$ for energies $W=1.525-1.925$~GeV~\cite{NSTAR11}. Curves: 
      $\eta$-MAID (dotted line), Bonn-Gatchina PWA (dashed line), and
      SAID (solid line).}
    \label{fig:brian}
  \end{center}
\end{figure}
 
\begin{figure}[tb]
  \begin{center}
    % please do not add file name extension this makes switching between latex and pdflatex easier
    \begin{tabular}{cc}
      \epsfig{file=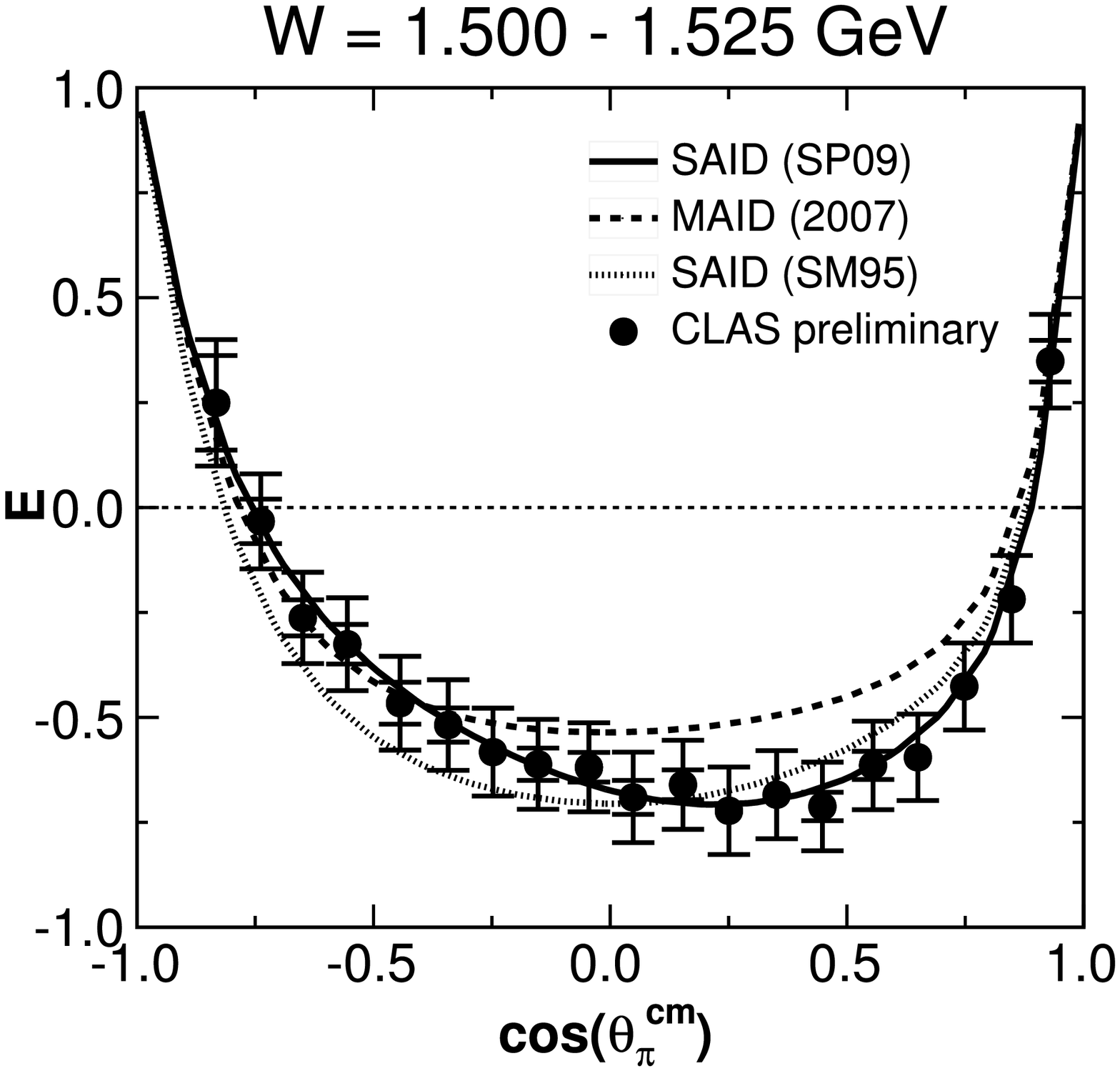,width=0.44\textwidth} &
      \epsfig{file=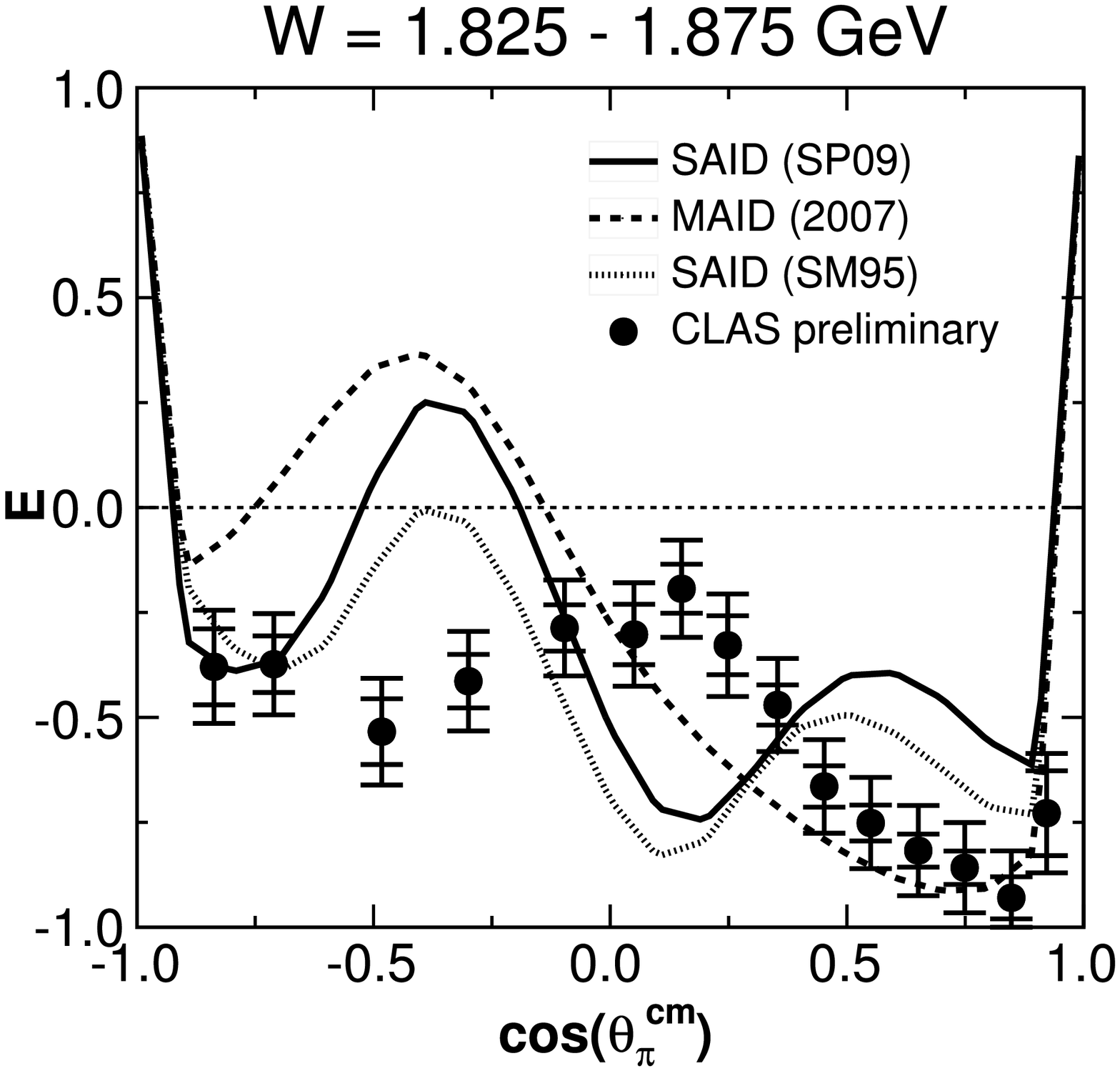,width=0.44\textwidth}
    \end{tabular}      
    \caption{Preliminary results of the double-pololarization observable $E$ 
      (helicity difference) for $\vec{\gamma} \vec{p}\to n\pi^+$~\cite{NSTAR11}. The inner 
      error bars indicate stat. uncertainties; the outer error bars include a 10\,\% 
      sys. uncertainty, which is expected to be reduced in the final analysis. 
      The curves show solutions of the SAID SP09~\cite{Dugger:2009pn}, 
      MAID~\cite{Drechsel:1998hk} and SAID~SM95~PWA.}
    \label{fig:steffen}
  \end{center}
\end{figure}

Of particular importance are well-chosen decay channels that can help
isolate contributions from individual excited states and clarify their
importance. Photoproduction of $\eta$~mesons offers the distinct
advantage of serving as an {\it isospin filter} for the spectrum of
nucleon resonances and, thus, simplifies data interpretations and 
theoretical efforts to predict the excited states contributing to these 
reactions. Since the $\eta$~mesons have isospin $I=0$, the $N\eta$ 
final states can only originate (in one-step processes) from intermediate 
$I=1/2$ nucleon resonances.

The polarized cross section for the reaction $\vec{\gamma} \vec{p}\to 
p\eta$ of circularly-polarized photons on longitudinally-polarized
protons is given by:
\begin{eqnarray}\label{observable:E}
\frac{d\sigma}{d\Omega} = \frac{d\sigma}{d\Omega}_0\,(1-\Lambda_z\,\delta_{\,\odot}\,E)\,,
\end{eqnarray}
where $d\sigma/d\Omega_{\,0}$ is the unpolarized cross section. $\Lambda_z$ and
$\delta_{\,\odot}$ are the degrees of target and beam polarization, respectively. 
$E$ denotes the helicity difference.

Preliminary results of the helicity difference $E$ for $\vec{\gamma}\vec{p}
\to p\eta$ are shown in Fig.~\ref{fig:brian}. Since the $\eta$-threshold
is dominated by the $N(1535)S_{11}$ resonance, the observable exhibits
values close to unity for $W < 1.6$~GeV/$c^2$. The preliminary results
indicate that the observable remains positive below about $W = 2$~GeV,
shedding further light on contributing resonances.

\section{The Helicity Asymmetry $E$ for the Reaction $\vec{\gamma}\vec{p}\to n\pi^+$}
Although many of the {\it unobserved} baryon resonances may have small
couplings to $\pi N$, it is still important to study pion photoproduction. 
Polarization observables will help sift the several competing descriptions 
of the spectrum by more conclusively indicating which resonances are involved 
in elastic pion-nucleon scattering, as well as providing evidence for previously 
unidentified resonances. New resonances found in reactions like $\gamma 
N\to\pi N$ are expected to have masses larger than about 1.8~GeV/$c^2$, 
although the higher-mass resonance contributions are expected to be more 
important in double-meson photoproduction.

The current database for pion photoproduction is mainly populated by
unpolarized cross section data and single-spin observables. Fig.~\ref{fig:steffen}
shows preliminary results of the double-polarization $E$ for $\vec{\gamma}
\vec{p}\to n\pi^+$ (Eqn.~\ref{observable:E}). While the predictions shown 
in the figure agree nicely with the new data at low energies (left side),
discrepancies emerge at higher energies (right side) for $W\geq 1.7$
GeV/$c^2$. Single-pion photoproduction appears less well understood
than previously expected. For this reason, the present data will greatly
reduce model-dependent uncertainties.

\section{Polarization Observables for $\pi^+\pi^-$ Production 
  on the Proton}

One of the key experiments in the search for yet unobserved states is
the investigation of double-pion photoproduction. Quark models predict
large couplings of those states to $\Delta\pi$, for instance. The
five-dimensional cross section for the photoproduction of two
pseudoscalar mesons using longitudinal target polarization and
circularly-polarized (or unpolarized) beam can be written in the
form~\cite{Roberts:2004mn}:
\begin{eqnarray*}\label{TwoMesonEquation}
       I\, =\, I_{\,0}\,\{\,(\, 1\,+\,\Lambda_z\cdot P_z\,)\,
          +\, \delta_{\,\odot}\, (I^{\,\odot}\,
          +\,\Lambda_z\cdot P^{\,\odot}_z\, )\,\}\,,
\end{eqnarray*}
\noindent
where $I_{\,0}$ denotes the unpolarized cross section and
$\delta_{\,\odot}$ and $\Lambda_z$ denote the degree of beam and target
polarization, respectively. The additional polarization observables,
$P_z$ and $I^{\,\odot}$, for the two-meson final state arise since
the reaction is no longer restricted to a single plane. Fig.~\ref{fig:sung} 
shows an example for the observable $P_z$ in $\gamma \vec{p}\to p\pi^+\pi^-$
\cite{NSTAR11}. The variables $\phi$ and $\theta$ denote the
azimuthal and polar angle of the $\pi^+$ in the rest frame of the 
two mesons. The observable acquires surprisingly large values for 
cos\,$\theta_{\pi^{+}} > 0$ with the statistical errors in some cases smaller 
than the symbol size. The expected odd behavior of the distribution 
is clearly visible. 

\begin{figure}[t]
  \begin{center}
    % please do not add file name extension this makes switching between latex and pdflatex easier
    %\includegraphics[width=1.0\textwidth]{SUNGKYUN/Ave_Pz_energyIndex05}
    \epsfig{file=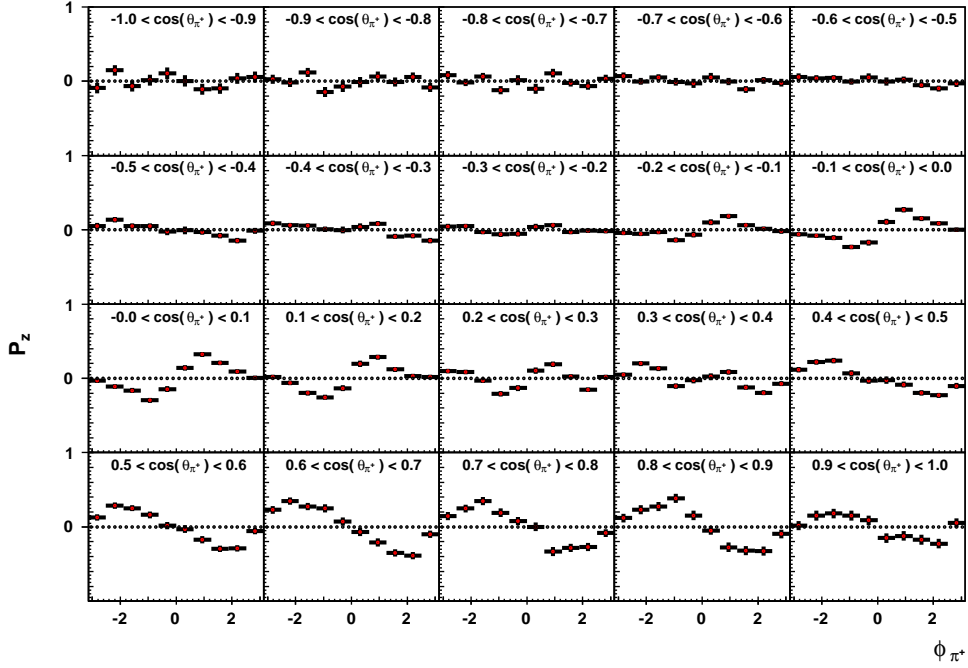,width=0.9\textwidth}
    \caption{Preliminary target asymmetry, $P_z$, from FROST for the
      reaction $\gamma \vec{p}\to p\pi^+\pi^-$ for $E_\gamma \in
      [0.7,\,0.8]$~GeV~\cite{NSTAR11}. Error bars are statistical only.}
    \label{fig:sung}
  \end{center}
\end{figure}

\section{Summary and Conclusion}

The goal of measuring a sufficient number of polarization
observables to unambiguously construct the scattering
amplitude for a given channel is within reach. New resonance 
candidates have been proposed on the basis of recent 
high-quality photoproduction data, though a clear pattern of
new states has not yet emerged. These efforts will soon shed light on
the open questions concerning the spectrum of baryon resonances. A
better understanding of QCD and the phenomenon of confinement appears
on the horizon.

%%%%%%%%%%%%%%%%%%%%%%%%%%%%%%%%%%%%%%%%%%%%%%%%%%%%%%%%%%%%%%%%%%%%%%%%%%%%%%%%%
% acknowledgements (optional)
\acknowledgements{%
Work supported in parts by the U.S. Department of Energy:
DE-FG02-92ER40735. Jefferson Science Associates operates the Thomas
Jefferson National Accelerator Facility under DOE contract
DE-AC05-06OR23177.
}

%%%%%%%%%%%%%%%%%%%%%%%%%%%%%%%%%%%%%%%%%%%%%%%%%%%%%%%%%%%%%%%%%%%%%%%%%%%%%%%%%
% bibliographic items can be constructed using the LaTeX format in SPIRES
% see http://www.slac.stanford.edu/spires/hep/latex.html
% SPIRES will also supply the CITATION line information; please include it

%
%%%%%%%%%%%%%%%%%%%%%%%%%%%%%%%%%%%%%%%%%%%%%%%%%%%%%%%%%%%%%%%%%%%%%%%%%%%%%%%%%

}  % do not remove

%%% Local Variables: 
%%% mode: latex
%%% TeX-master: "../hadron2011.tex"
%%% End: 


\begin{thebibliography}{99}

\bibitem{Dugger:2009pn}
  M.~Dugger {\it et al.} [CLAS Collaboration],
  %``pi+ photoproduction on the proton for photon energies from 0.725 to 2.875-GeV,''
  Phys.\ Rev.\  {\bf C79}, 065206 (2009);
  %[arXiv:0903.1110 [hep-ex]].

\bibitem{Williams:2009rc}
  M.~Williams {\it et al.} [ CLAS Collaboration ],
  %``Partial wave analysis of the reaction gamma p ---> p omega and the search for nucleon resonances,''
  Phys.\ Rev.\  {\bf C80}, 065209 (2009).
  %[arXiv:0908.2911 [nucl-ex]].

\bibitem{Kamano:2010ud}
  H.~Kamano, S.~X.~Nakamura, T.~-S.~H.~Lee, T.~Sato,
  %``Extraction of P11 resonances from pi N data,''
  Phys.\ Rev.\  {\bf C81}, 065207 (2010).
  %[arXiv:1001.5083 [nucl-th]].

\bibitem{Chiang:1996em}
  W.~-T.~Chiang, F.~Tabakin,
  %``Completeness rules for spin observables in pseudoscalar meson photoproduction,''
  Phys.\ Rev.\  {\bf C55}, 2054-2066 (1997).
  %[nucl-th/9611053].

\bibitem{Bradford} R.~Bradford et al., Phys.\ Rev.\ {\bf C73}, 035202 (2006);
  Phys.\ Rev.\ {\bf C75}, 035205 (2007).
 
%\cite{Hleiqawi:2007ad}
\bibitem{Hleiqawi:2007ad}
  I.~Hleiqawi {\it et al.} [ CLAS Collaboration ],
  %``Cross-sections for the gamma p ---> K*0 Sigma+ reaction at E(gamma) = 1.7-GeV - 3.0-GeV,''
  Phys.\ Rev.\  {\bf C75}, 042201 (2007);\\ Erratum-ibid.C76:039905,2007.
  %[nucl-ex/0701036].

\bibitem{McCracken:2009ra}
  M.~E.~McCracken {\it et al.} [ CLAS Collaboration ],
  %``Differential cross section and recoil polarization measurements for the gamma p to K+ Lambda reaction using CLAS at Jefferson Lab,''
  Phys.\ Rev.\  {\bf C81}, 025201 (2010).
  %[arXiv:0912.4274 [nucl-ex]].

%\cite{AnefalosPereira:2009zw}
\bibitem{AnefalosPereira:2009zw}
  S.~A.~Pereira {\it et al.} [ CLAS Collaboration ],
  %``Differential cross section of gamma n to K+ Sigma- on bound neutrons with incident photons from 1.1 to 3.6 GeV,''
  Phys.\ Lett.\  {\bf B688}, 289-293 (2010).
  %[arXiv:0912.4833 [nucl-ex]].

\bibitem{Chen:2009sda}
  W.~Chen {\it et al.} [ CLAS Collaboration ],
  %W.~Chen, T.~Mibe, D.~Dutta, H.~Gao, J.~M.~Laget, M.~Mirazita, P.~Rossi, S.~Stepanyan {\it et al.},
  %``A Measurement of the differential cross section for the reaction gamma n ---> pi- p from deuterium,''
  Phys.\ Rev.\ Lett.\  {\bf 103}, 012301 (2009).
  %[arXiv:0903.1260 [nucl-ex]].

\bibitem{Mecking:2003zu}
  B.~A.~Mecking {\it et al.} [ CLAS Collaboration ],
  %``The CEBAF Large Acceptance Spectrometer (CLAS),''
  Nucl.\ Instrum.\ Meth.\  {\bf A503}, 513-553 (2003).

\bibitem{Sober:2000we}
  %D.~I.~Sober, H.~Crannell, A.~Longhi, S.~K.~Matthews, J.~T.~O'Brien, B.~L.~Berman, W.~J.~Briscoe, P.~L.~Cole {\it et al.},
  D.~I.~Sober {\it et al.},
  %``The bremsstrahlung tagged photon beam in Hall B at JLab,''
  Nucl.\ Instrum.\ Meth.\  {\bf A440}, 263-284 (2000).

\bibitem{NSTAR11}
  B.~Morrison, S.~Park, NSTAR 2011, Jefferson Lab;
  S.~Strauch, arXiv:1108.3050 [nucl-ex].

%\cite{Drechsel:1998hk}
\bibitem{Drechsel:1998hk}
  D.~Drechsel, O.~Hanstein, S.~S.~Kamalov, L.~Tiator,
  %``A Unitary isobar model for pion photoproduction and electroproduction on the proton up to 1-GeV,''
  Nucl.\ Phys.\  {\bf A645}, 145-174 (1999).
  %[nucl-th/9807001].

%\cite{Roberts:2004mn}
\bibitem{Roberts:2004mn}
  W.~Roberts, T.~Oed,
  %``Polarization observables for two-pion production off the nucleon,''
  Phys.\ Rev.\  {\bf C71}, 055201 (2005).
  %[nucl-th/0410012].

\end{thebibliography}
\end{document}